\documentclass[10pt,journal]{IEEEtran}

\usepackage{cite}
\usepackage{graphicx}
\usepackage{algorithm}
\usepackage{algorithmic}
\usepackage[hyphens]{url}
\usepackage{amsmath}
\usepackage{wrapfig}

\begin{document}

\title{Pixelated VLC-backscattering for Self-charging Indoor IoT Devices}

\author{Sihua Shao, Abdallah Khreishah, Hany Elgala
\thanks{Sihua Shao and Abdallah Khreishah are with the Department of Electrical and Computer Engineering,
New Jersey Institute of Technology, email: ss2536@njit.edu, abdallah@njit.edu}
\thanks{Hany Elgala is with the Department of Computer Engineering,
University at Albany, email: helgala@albany.edu}\vspace{-30pt}}

\maketitle

\begin{abstract}
Visible light communication (VLC) backscatter has been proposed as a wireless access option for Internet of Things (IoT). However, the throughput of the state-of-the-art VLC backscatter is limited by simple single-carrier pulsed modulation scheme, such as on-off keying (OOK). In this paper, a novel pixelated VLC backscatter is proposed and implemented to overcome the channel capacity limitation. In particular, multiple smaller VLC backscatters, switching on or off, are integrated to generate multi-level signals, which enables the usage of more advanced modulation schemes than OOK. Based on experimental results, rate adaptation at different communication distances can be employed to enhance the achievable data rate. Compared to OOK, the data rate can be tripled when 8-PAM is used at 2 meters. In general, $n$-fold throughput enhancement is realized by utilizing $n$ smaller VLC backscatters while incurring negligible additional energy using the same device space as that of a single large backscatter.
\end{abstract}
\vspace{-15pt}
\section{Introduction}\label{sec1}
It is expected that by 2020, the Internet will consist of 50 billion devices \cite{cisco2016iot}, which leads to imperative design of the Internet of Things (IoT). The IoT should be able to link every small object to the Internet and to enable an exchange of data never available before. However, to braze the trail for IoT, several challenges need to be resolved. Connecting all these devices to the Internet through wire cable is impractical due to the deployment of wires, complexity added to the small IoT devices, and mobility of the devices. Thus the IoT devices are expected to be connected to Internet via wireless medium. Nevertheless, with the increase in the extremely large amount of wireless access devices, these IoT devices will compete with the users' devices on allocating the spectrum and the ``spectrum crunch" \cite{hanchard2010fcc} problem will be exacerbated if there is no innovative solution to significantly enhance the spectral efficiency. Furthermore, since it is impractical and cost-inefficient to replace the batteries of all these devices or powering them through power cables, self-sustained operation enabled by energy harvesting needs to be tackled. The energy harvested from external sources (e.g. solar power, thermal energy and kinetic energy) is sufficient for the signal generation of the IoT devices \cite{priya2009energy}. However, what can be harvested by IoT devices (in the order of 100 $\mu$Watts) restricts the distance that transmitted signals can travel. Therefore, connecting to cellular networks or even WiFi is not usually a viable option. This distance constraint results in the requirement of huge infrastructure support, like access points and backhaul links, in order to achieve a small distance reuse factor. Also, trying to achieve such a small reuse factor causes the so-called ``backhaul challenge" \cite{chia2009next}.

One proposed solution to the above challenges is to use RF-backscattering \cite{kellogg2014wi,bharadia2015backfi}. Small IoT devices harvest energy from the ambient RF signals broadcasted by TV towers \cite{liu2013ambient} or WiFi signals generated by wireless routers, and modulate the reflected RF signals by varying the antenna's impedance, which affects the amount of signal that is reflected by the RF backscatter. While RF-backscattering provides an option for the Internet access of IoT devices, it has several inherent drawbacks. First, the uplink data rate depends on the amount of downlink traffic. For instance, in \cite{kellogg2014wi}, in order to achieve an uplink data rate of 1 kbps, the WiFi router has to send at a data rate of more than 1.5 Mbps. In \cite{bharadia2015backfi}, it needs the downlink traffic at 24 Mbps in a range of 5 meters, to support the uplink data rate of 1 Mbps. Also, due to the omnidirectional propagation of the RF backscatter signals, interference among the RF backscatters and the uplink data traffic from the users' devices will be inevitable and destructive, especially when the number of IoT devices is large. Furthermore, since the maximum communication distance is short (e.g. 2-5 meters reported in \cite{kellogg2014wi,bharadia2015backfi}), at least one access point needs to exist within this distance. This fact leads to the expensive infrastructure construction cost. One solution to this problem could be placing a reader \cite{kellogg2014wi}, such as a mobile phone, within the maximum communication distance, which performs a signal relaying functionality between the access point and the IoT devices. However, this creates additional traffic from the reader, which competes with the mobile traffic and exacerbates the ``spectrum crunch". Also, these readers might not exist all of the time and using mobile phones as readers might cause privacy problems. Another limitation of the RF-backscattering is that we might not be able to use it in a limited RF environment, such as a hospital and an airplane.

\begin{figure*}
  \centering
  \begin{minipage}{.36\linewidth}
    \includegraphics[width=1.0\textwidth]{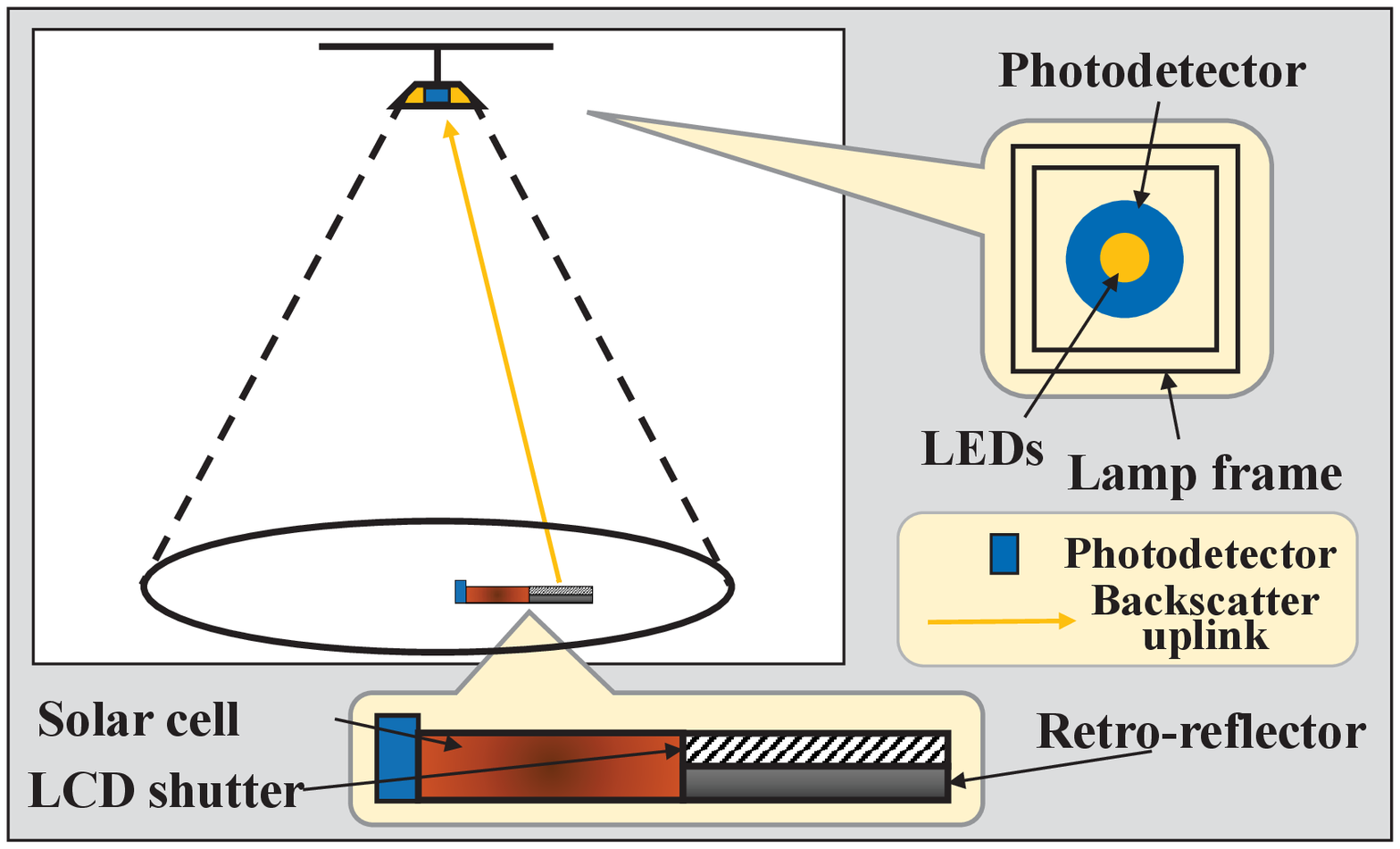}
  \vspace{-10pt}
  \caption{VLC backscatter system architecture}
  \vspace{-15pt}
  \label{fig_architecture}
  \end{minipage}
  \begin{minipage}{.30\linewidth}
   \includegraphics[width=1.0\textwidth]{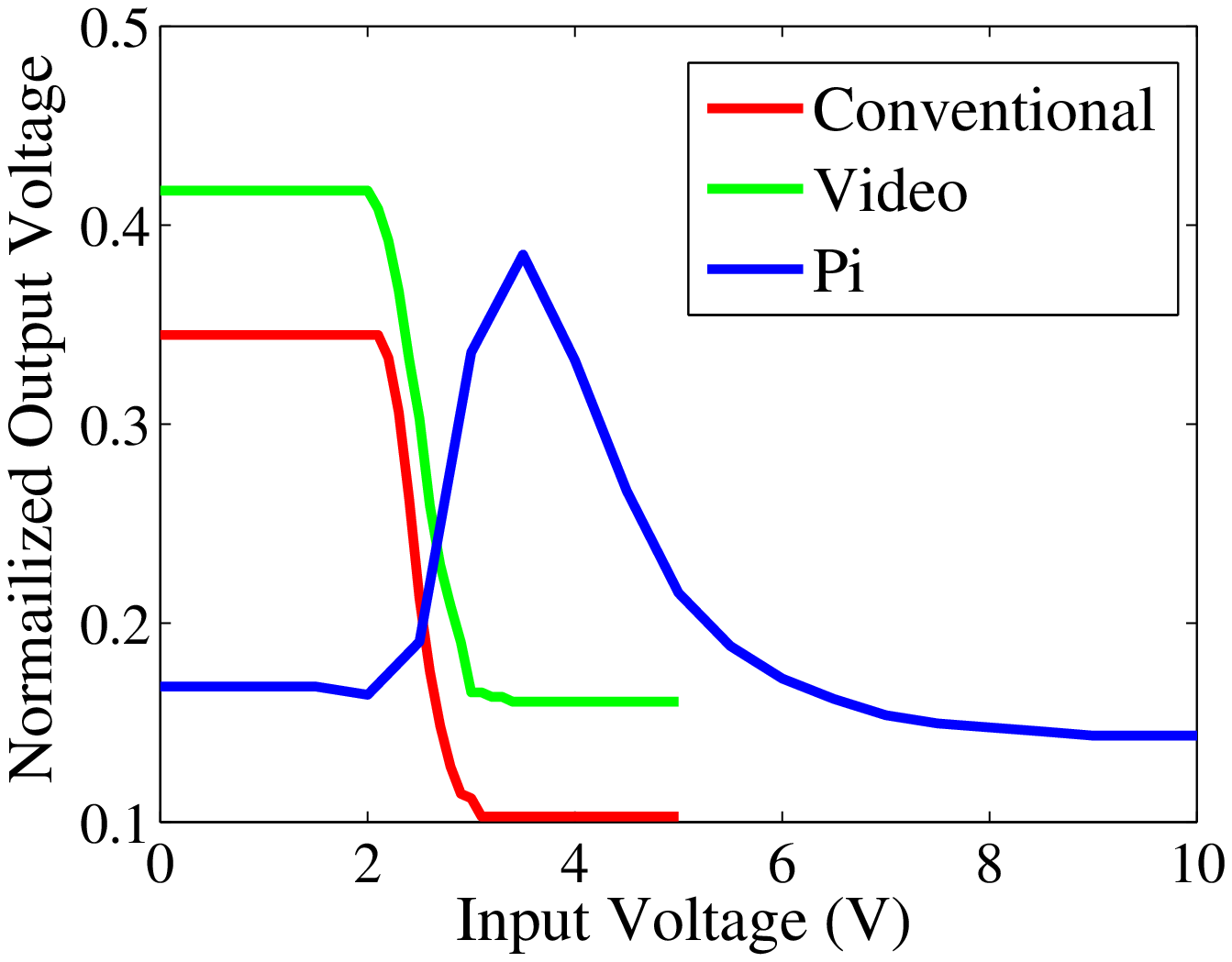}
  \vspace{-10pt}
  \caption{The IO curves for three different types of LCD shutters}
  \vspace{-15pt}
  \label{fig_IO_voltage}
  \end{minipage}
  \begin{minipage}{.30\linewidth}
    \includegraphics[width=1.0\textwidth]{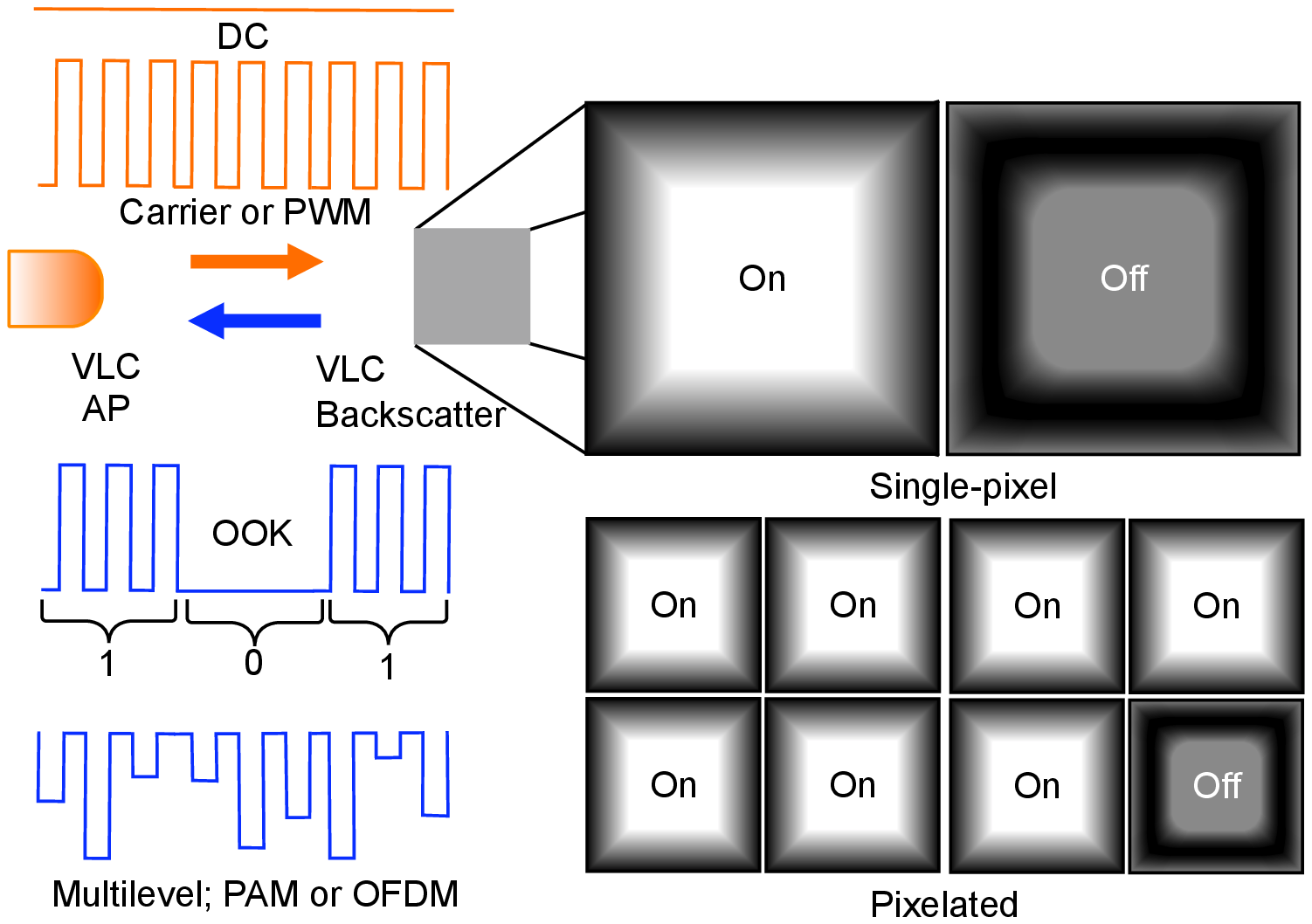}
  \vspace{-10pt}
  \caption{The proposed pixelated based system}
  \vspace{-15pt}
  \label{fig_Pixelated_concept}
  \end{minipage}
\end{figure*}

Due to the above-mentioned problems, another solution is introduced as the visible light communications (VLC) backscattering \cite{mansell2002modulated,komine2003bidirectional,rosenkrantz2014modulated,li2015retro}. The VLC backscatter harvests energy from the existing indoor lighting infrastructure and performs modulation on the reflected light beam as long as the illumination is available. As shown in Fig.~\ref{fig_architecture}, a small solar cell is used to harvest energy and a liquid crystal display (LCD) shutter, switching on or off, is used to modulated the reflected optical signals from a retro-reflector. This is motivated by the fact that whenever communication is needed illumination is also needed most of the time. Compared to RF backscatter, VLC backscatter solves the backhaul challenge by utilizing the existing infrastructure with slight modification \cite{shao2014indoor,shao2015design,ayyash2015coexistence} (i.e. adding driver and photodetector to the light source and utilizing the power line communication or power over Ethernet). Backscattering light beam is also directional, which mitigates the interference problem. VLC backscatter also does not require a reader to relay the uplink data transmission. A shutter modulating the reflected light is proposed in \cite{mansell2002modulated} and the corner cube reflector maintaining the directionality of reflected beam is presented in \cite{komine2003bidirectional,rosenkrantz2014modulated}. The state-of-the-art practical work that brings VLC backscatter into the IoT field is \cite{li2015retro}. In \cite{li2015retro}, a VLC backscatter prototype is presented, which uses a solar cell to harvest the optical energy from indoor light sources and modulate the reflected light by a LCD shutter. Nevertheless, no VLC backscatter system can leverage a modulation scheme that is more advanced than on-off keying (OOK). This is because the nonlinearity polarization of LCD shutters. Three types of LCD shutters from Liquid Crystal Technologies \cite{LCT}, named conventional, video and Pi shutters, are evaluated by measuring the transparency with variable input direct current (DC) voltage. As shown in Fig.~\ref{fig_IO_voltage}, the output amplitude drops suddenly when the input voltage reaches a certain level. Therefore, applying advanced modulation schemes, such as pulse amplitude modulation (PAM) or orthogonal frequency division multiplexing (OFDM), is not available when only one VLC backscatter is used, which is different from the impedance matching approach applied in RF backscatter \cite{bharadia2015backfi}.



In this work, we propose and implement a novel VLC backscatter, called pixelated VLC backscatter, which uses multiple smaller reflectors and LCD shutters to form numbers of pixels. Each pixel can switch on or off independently in order to produce multi-level signals. As shown in Fig.~\ref{fig_Pixelated_concept}, with the same size of single pixel, pixelated backscatter increases the number of pixels by reducing the size of each pixel. The impact of pixelated design on the bit error rate (BER) performance will be evaluated in the experiment section. To the best of our knowledge, this pixelated backscattering design is {\it allowing for the first time more advanced modulation schemes than OOK} for small IoT devices using optical backscattering. Based on experiments, it is observed that the energy consumption of the pixelated design causes negligible overhead, which validates the availability of the pixelated VLC backscater under most indoor environments. Based on our testbed, the throughput achieved by the pixelated VLC backscatter is 600 bps at 2 meters, which is highly restricted by the response time of the off-the-shelf LCD shutters (5 ms). By reducing the size of LCD shutter, the capacitance of the LCD shutter is expected to decrease. Due to the RC time constant \cite{ahn2000method}, smaller capacitance implies lower charging delay, which is the time consumed by a LCD shutter to tune its polarization. Therefore, reducing the size of LCD shutters can potentially enhance the throughput. Here, we do not customize the size of LCD shutters since we use the off-the-shelf samples.

\vspace{-5pt}
\section{Theoretical Analysis}\label{sec2}
In this section, we theoretically evaluate the benefit of adding pixels and the relationship between the maximum communication distance and the target BER when certain number of pixels is used and PAM is considered.

\begin{wrapfigure}{r}{0.21\textwidth}
  \vspace{-6pt}
  \begin{center}
    \includegraphics[width=0.21\textwidth]{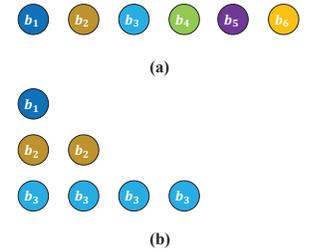}
  \end{center}
  \vspace{-10pt}
  \caption{Structure of pixels: (a) pixels with the same size; (b) binary-weighted clustering pixels}
  \vspace{-6pt}
  \label{fig_pixelated_theoretical}
\end{wrapfigure}
In order to control the pixelated VLC backscatter while achieving a target backscattered signal quality, our main goal is to investigate the trade-off between the required number and size of pixels, signal-to-noise ratio (SNR) and resolution. Assume that the array, consists of identical pixels, i.e. when switched on, each pixel backscatters the same amount of optical power proportional to incident optical flux on it, then the backscattered light is proportional to the number of pixels that are switched on (Fig.~\ref{fig_pixelated_theoretical} (a)). In this case, $2^m$ optical levels can be achieved by switching $2^m$ pixels. To reduce the number of pixels with a given resolution, we can try to use a binary-weighted structure (Fig.~\ref{fig_pixelated_theoretical} (b)). In this case, the $m^{th}$ cluster contains $2^{m-1}$ same-size pixels. In the following context, we refer to $m$ instead of $2^{m-1}$ when we discuss the number of pixels, since the cluster of same-size pixels can be viewed as different-size pixels.

Here, PAM is considered as the modulation scheme. The $2^m$ optical levels enhance the bit rate by $m$ times, when compared to the OOK, of which the bit rate is 1 bit/symbol. Given the symbol rate and assume the target BER is always satisfied, the throughput of $m$ pixles based pixelated VLC backscatter is $m$ times that of the single pixel. Since the IoT device is conceptually very small, assuming a certain area of VLC backscatter, we investigate the trade-off between the number of pixels and communication distance for a target BER.

Consider the M-PAM, the required number of pixels is $log_2{M}$. The relationship between BER and SNR is \cite{hranilovic2006wireless}
\begin{align}
&\text{BER}=\frac{2(M-1)}{Mlog_2M}Q(\frac{1}{M-1}\sqrt{SNR})
\end{align}
where SNR denotes the square of the ratio of peak-to-peak amplitude and noise. Here we consider the noise as the worst case, which is the case that all the pixels are switched on. Since the signal strength attenuates by a factor of $\frac{1}{d^2}$, where $d$ denotes the communication distance, the relationship between BER and distance $d$ can be written as
\begin{align}
&\text{BER}=\frac{2(M-1)}{Mlog_2M}Q(\frac{1}{M-1}\sqrt{f(\frac{1}{d^2})})\nonumber
\end{align}
It can be easily found that, given the BER, the number of pixels $log_2{M}$ is inversely proportional to the maximum communication distance $d$.
\vspace{-10pt}
\section{Experimental Results}\label{sec3}
\subsection{Testbed setup}
The testbed consists of an 8.5 Watts white LED bulb, a microcontroller MSP430G2553 from Texas Instrument, 3 LCD shutters from Liquid Crystal Technologies, 3 super base ball stages, 3 cage mounted irises with diameter 20.0 mm and Photodetector PDA 36A from Thorlabs, 3 small mirrors, oscilloscope Tektronix MDO-4034, and a DC power supply.

\begin{figure}
\centering
\includegraphics[width=0.4\textwidth]{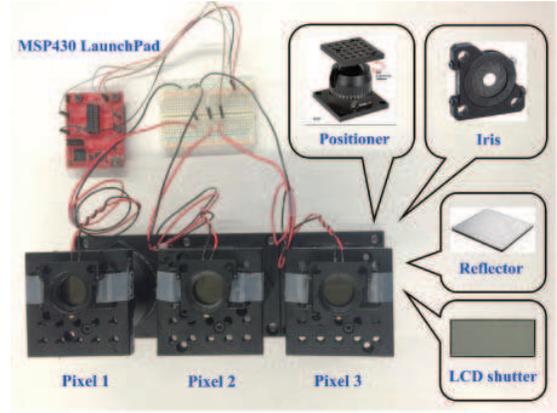}
\vspace{-5pt}
\caption{Testbed of pixelated VLC backscatter}
\vspace{-15pt}
\label{fig_testbed}
\end{figure}

The response time of the LCD shutters is around 5 ms, thus the maximum modulation frequency is set at 200 Hz. In \cite{li2015retro}, the response time of their LCD shutter is 2 ms, thus the effective bandwidth of their LCD shutter is larger than that of ours. The super base ball stages are used to control the azimuth and elevation angle of the pixels in order to perform the alignment. Since this testbed works as a proof-of-concept to investigate the performance of pixlated VLC backscatter, manually controlled positioners are used for orientation. Nevertheless, in production, the orientation process can be performed by piezoelectric actuators \cite{dosch1992self} which allow the reflectors to change their orientation with low amount of energy. The cage mounted irises are installed upon the LCD shutters to manipulate the area of each pixel. For instance, when one pixel is activated, the diameter is set at 20.0 mm; when two pixels are activated, the diameter of the 1st pixel is set at 16.33 mm, and the diameter of the 2nd pixel is set at 11.54 mm, to keep the total reflector area the same as that of one pixel. Note that by customizing the size of each pixel instead of controlling the reflector area through iris, the effective bandwidth of each pixel could be enhanced (i.e. above 200 Hz). Here, we use the off-the-shelf samples instead of customizing the size of the LCD shutters. The transmitted signals are pre-compiled in MSP430G2553 and this microcontroller is powered by a DC power supply with 3 V. The optical signals modulated by LCD shutters are captured by the photodetector and the waveform is displayed on the oscilloscope for further analysis. The VLC backscatter prototype, including the microcontroller MSP430 and three pixels, is shown in Fig.~\ref{fig_testbed}. For each pixel, a LCD shutter is placed upon a reflector (i.e. mirrors), and a circular iris is fixed upon the LCD shutter to control the effective backscatter area.

\begin{table}
\parbox{.45\linewidth}{
\centering
\caption{DCO operates at 3 V}
\begin{tabular}{ccc}
\hline
\hline
Frequency&Current\\
10 Hz&337 $\mu$A\\
100 Hz&341 $\mu$A\\
500 Hz&354 $\mu$A\\
\hline
\hline
\vspace{-20pt}
\end{tabular}\label{MSP430_DCO_3V}
}
\hfill
\parbox{.45\linewidth}{
\centering
\caption{VLO operates at 3 V}
\begin{tabular}{ccc}
\hline
\hline
Frequency&Current\\
10 Hz&68 $\mu$A\\
100 Hz&69 $\mu$A\\
500 Hz&70 $\mu$A\\
\hline
\hline
\vspace{-20pt}
\end{tabular}\label{MSP430_VLO_3V}
}
\end{table}

\begin{figure*}
  \centering
  \begin{minipage}{.32\linewidth}
  \centering
    \includegraphics[width=0.8\textwidth]{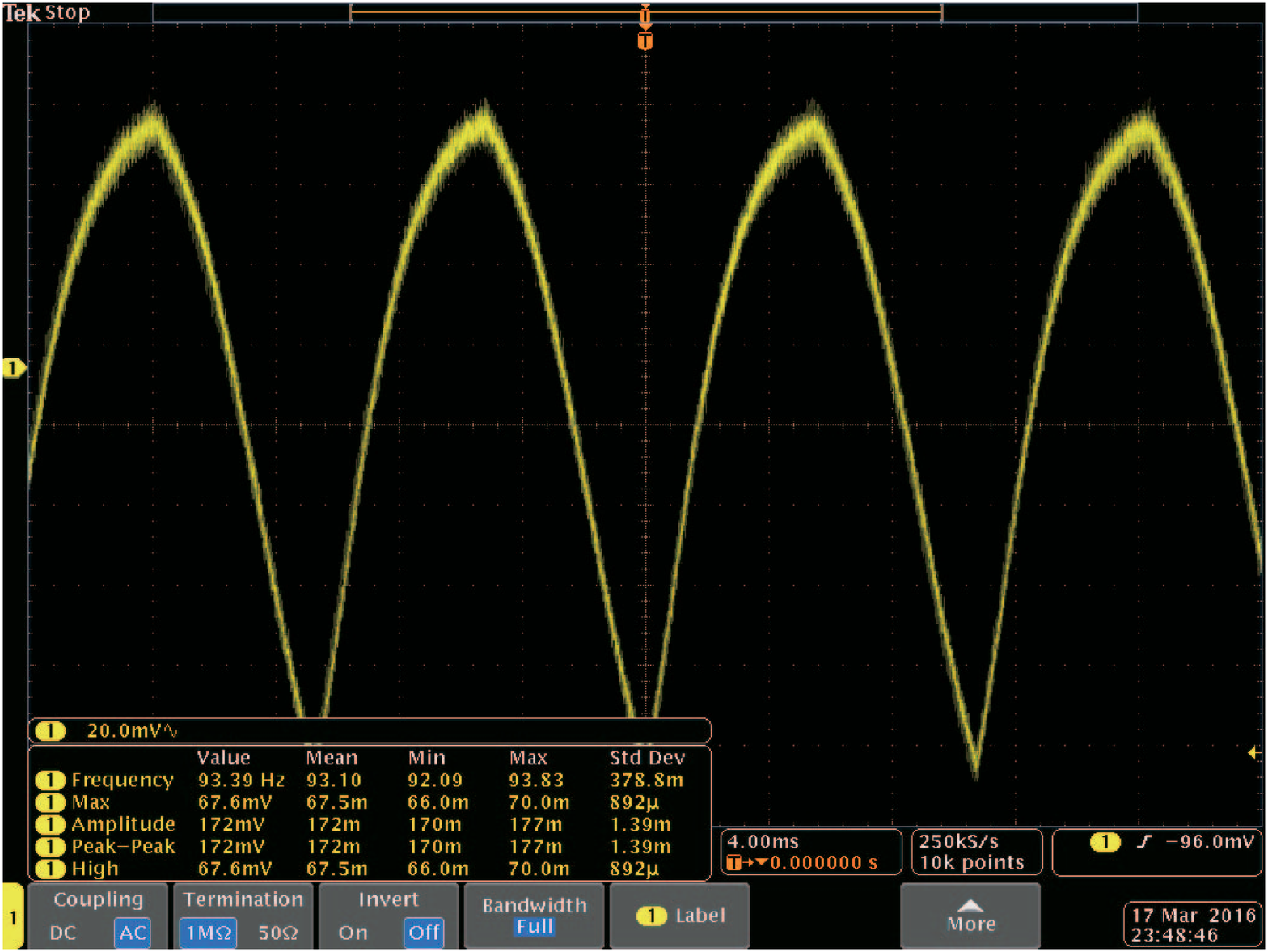}
  \caption{Single pixel transmits at the distance of 2 meters and generates 2 signal levels}
  \vspace{-10pt}
  \label{fig_2m_single_pixel}
  \end{minipage}
  \begin{minipage}{.32\linewidth}
  \centering
   \includegraphics[width=0.8\textwidth]{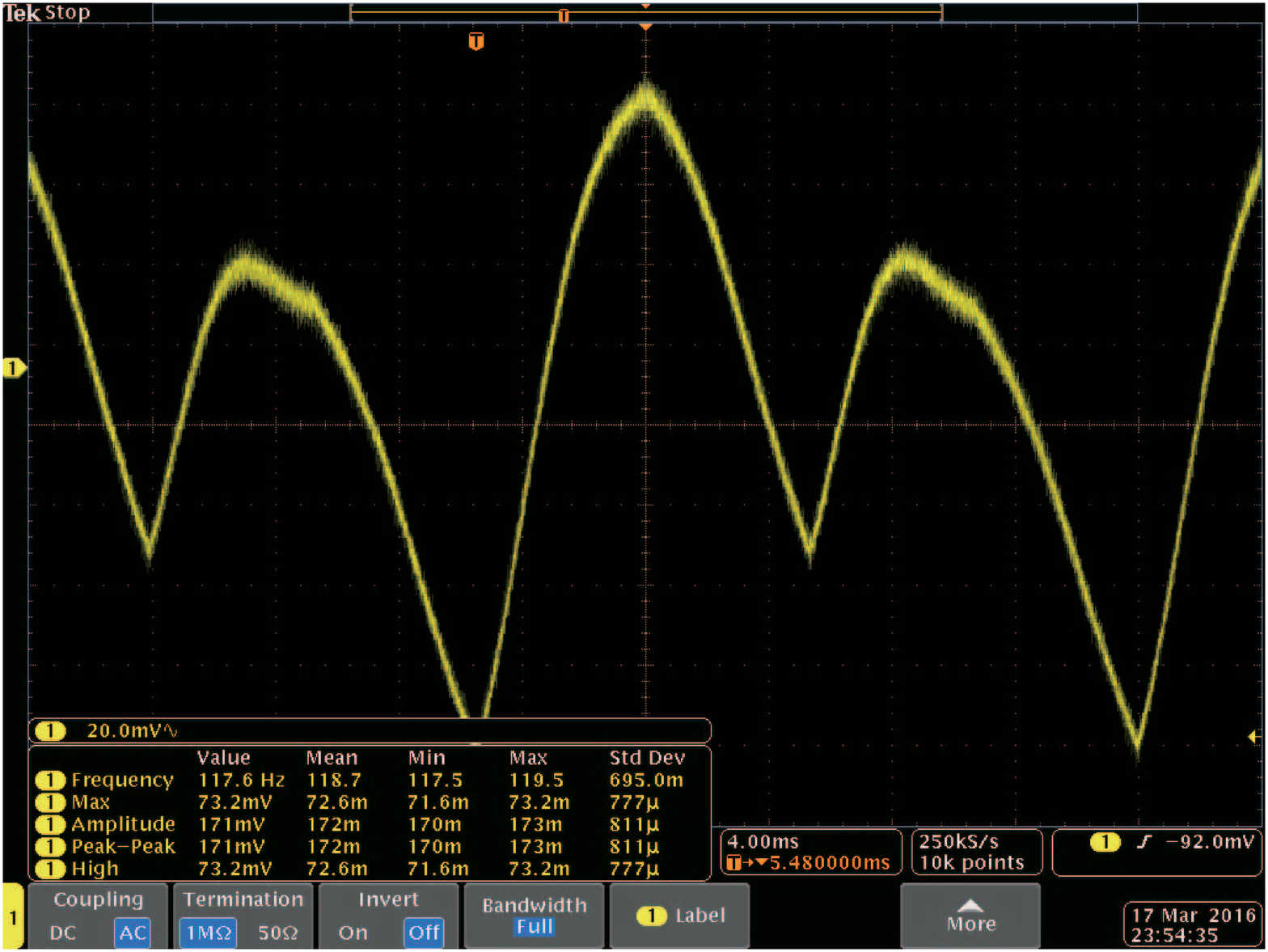}
  \caption{Two pixels transmit at the distance of 2 meters and generates 4 signal levels}
  \vspace{-10pt}
  \label{fig_2m_two_pixels}
  \end{minipage}
  \begin{minipage}{.32\linewidth}
  \centering
    \includegraphics[width=0.8\textwidth]{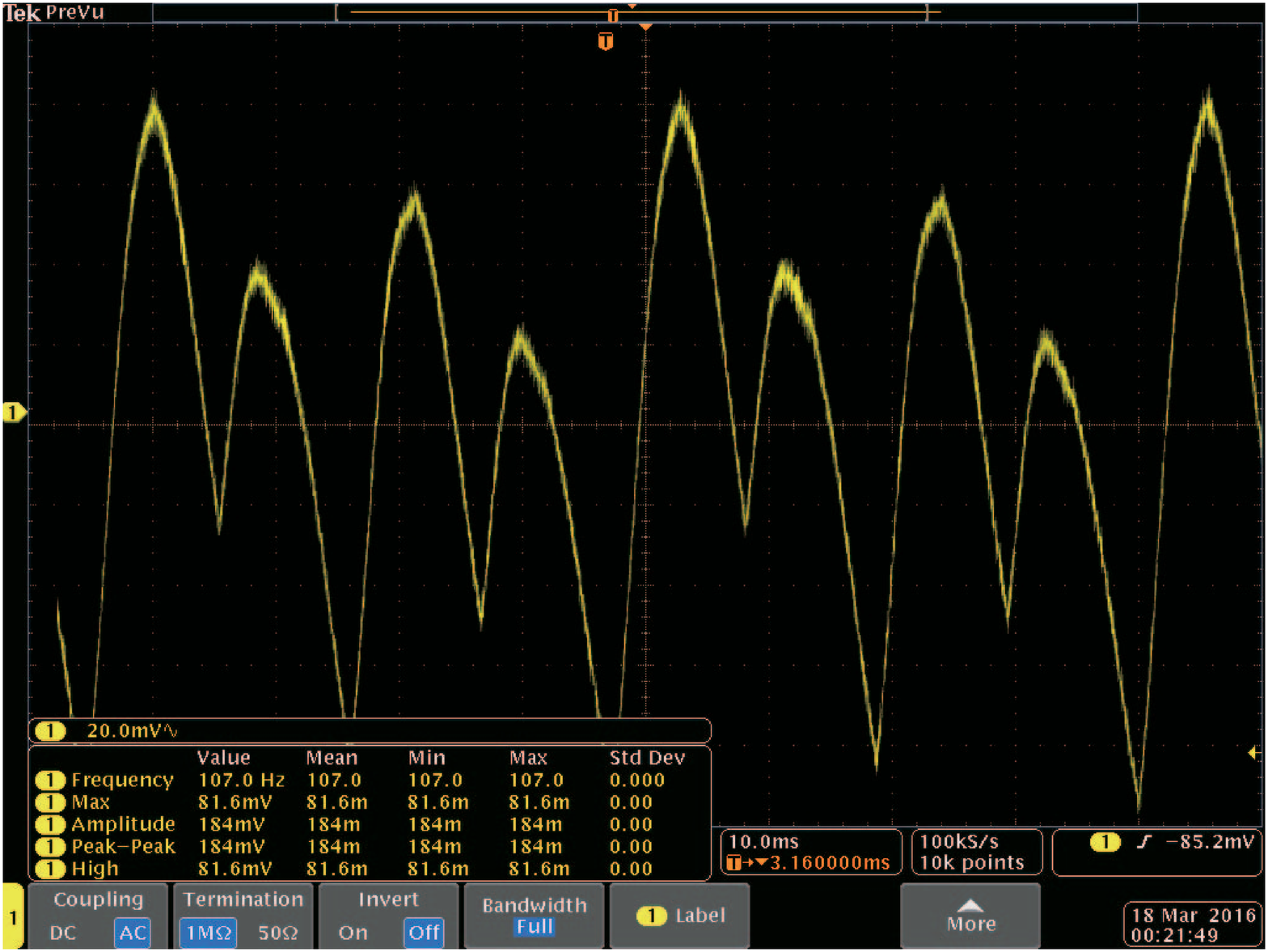}
  \caption{Three pixels transmit at the distance of 2 meters and generates 8 signal levels}
  \vspace{-10pt}
  \label{fig_2m_three_pixels}
  \end{minipage}
\end{figure*}
\vspace{-10pt}
\subsection{Power consumption}
First, we measure the power consumption of microcontroller and the LCD shutter and we find that the microcontroller consumes power in the order of magnitude 100 $\mu$W while the LCD shutter consumes less than 0.2 $\mu$W at 200 Hz (i.e. maximum modulation frequency). Therefore, the power consumption of the microcontroller is dominant in the VLC backscatter system.

The default clock module of the MSP430G2553 is an internal digitally controlled oscillator (DCO), which operates at 1 MHz. When the DC power supply provides 3 V, the current is around 340 $\mu$A at 100 Hz (Table~\ref{MSP430_DCO_3V}). The required power $\sim$1 mW is usually too high to be harvested from extreme indoor illumination ($\sim$200 lux) by a small solar cell (e.g. 50 mm $\times$ 50 mm). Since the maximum modulation frequency of the LCD shutter is only 200 Hz, the internal very-low-power low-frequency oscillator (VLO) is selected instead, of which the typical frequency is 12 kHz. With 3 V DC supply, the power consumption of the MSP430G2553 is reduced to around 200~$\mu$W (Table~\ref{MSP430_VLO_3V}) when VLO is chosen. By performing the following calculation: $P=E_{lux}\times I\times A\times E_{solar}$, where $E_{lux}$ denotes the conversion ratio from lux to watt/cm$^2$, $I$ denotes the illumination in unit lux, $A$ represents the solar cell area, and $E_{solar}$ represents the solar cell efficiency factor, we can find the electrical power $P$ that can be generated by solar cell under indoor environment. We have $E_{lux}=1.46\times10^{-7}~watt/cm^2/lux$ \cite{Lux}, $I=200~lux$, and $A=25~cm^2$. Based on the state-of-the-art solar cell technology, the solar cell efficiency is up to 40\% \cite{green2015solar}. Substituting these values into the solar power calculation, the output $P=292~\mu W$. It means that the amount of power consumed by microcontroller (i.e. around 200 $\mu W$) can be harvested from indoor lighting. Note that, as shown in Table~\ref{MSP430_DCO_3V} and Table~\ref{MSP430_VLO_3V}, the impact of changing the frequency on power consumption is negligible.
\vspace{-10pt}
\subsection{Throughput vs. distance}
Now we start evaluating the trade-off between achievable throughput and the maximum communication distance, as discussed in the Section \ref{sec2}. The number of pixels are varied from 1 to 3, corresponding to OOK, 4-PAM, and 8-PAM, respectively. The measurement distance is ranging from 2 meters to 5 meters. And the modulation frequency is set at 200 Hz, which means that the symbol rate is 200 symbols per second. The results for different number of pixels transmitting at 2 meters are shown in Fig.~\ref{fig_2m_single_pixel} to Fig.~\ref{fig_2m_three_pixels}. As it is observed from Fig.~\ref{fig_2m_three_pixels}, the Euclidean distance of the constellation between different signal levels is still much larger than the noise when 8-PAM is applied. In Fig.~\ref{fig_2m_single_pixel}, each signal level represents one bit and thus the achievable throughput is 200 bps. In Fig.~\ref{fig_2m_three_pixels}, each signal level represents three bits and thus the throughput is increased to 600 bps. Therefore, with a relatively short communication distance, the $M$ pixels based pixelated VLC backscatter is able to provide $M$ times of the throughput of the single pixel VLC backscatter with negligible energy overhead and the same reflector area.

%
%

To evaluate the trade-off between achievable throughput and the maximum communication distance, we measure the SNR at different distances and show the results in Table~\ref{Distance_vs_SNR}. Note that the SNR values measured at different distances are the same for OOK, 4-PAM, and 8-PAM. Based on equation (1), we calculate the required SNR for different modulation schemes when the target BER is $10^{-3}$. The results are shown in Table~\ref{Modulation_scheme_vs_SNR}. From these results, we observe that the OOK modulation scheme can achieve the target BER above 5 meters, while the 8-PAM can only achieve the target BER at around 2 meters. Therefore, we can conclude that, based on our testbed, 600 bps can be achieved at 2 m, 400 bps can be achieved at 3 m , and 200 bps can be achieved above 5 m. The higher order PAM is beneficial at shorter distance by doubling or even tripling the throughput, but has to sacrifice the maximum communication distance. This fact motivates performing the rate adaptation when pixelated VLC backscatter is applied. Modulation scheme with smaller Euclidean distance of the constellation is used at shorter communication distance to enhance the achievable throughput, while modulation scheme with larger Euclidean distance of the constellation is used at longer distance to guarantee the robust connectivity.

\begin{table}
\parbox{.35\linewidth}{
\centering
\caption{Distance vs. SNR}
\begin{tabular}{ccc}
\hline
\hline
2 m&3 m\\
26.55 dB&21.15 dB\\
\hline
\hline
4 m&5 m\\
18.80 dB&14.98 dB\\
\hline
\hline
\vspace{-20pt}
\end{tabular}\label{Distance_vs_SNR}
}
\hfill
\parbox{.55\linewidth}{
\centering
\caption{Modulation scheme vs. required SNR}
\begin{tabular}{ccc}
\hline
\hline
OOK & 4-PAM & 8-PAM\\
9.80 dB&19.10 dB&26.23 dB\\
200 bps&400 bps&600 bps\\
\hline
\hline
\vspace{-20pt}
\end{tabular}\label{Modulation_scheme_vs_SNR}
}
\end{table}
\vspace{-10pt}
\section{Conclusion}
In this work, a novel pixelated VLC backscatter system is proposed and implemented. With ultra-low power consumption overhead, $n$ pixels can enhance the throughput by $n$ times using the same reflector area when compared to single pixel VLC backscater with single carrier pulsed OOK. Based on our experimental results and using 8-PAM, 600 bps is achieved at 2 meters and this data rate can be still greatly enhanced if customizing the size of each pixel is available or the LCD shutter is replaced by a faster modulator. The worse BER performance at longer distance for higher order PAM motivates the rate adaptation of applying pixelated VLC backscatter. At short distance, higher order PAM is preferred to provide high throughput while at long distance, lower order PAM is needed to maintain the channel quality. As our future work, applying OFDM modulation scheme to pixelated VLC backscatter will be investigated.
\vspace{-10pt}
\bibliographystyle{IEEEtran}
\bibliography{draft_paper}

\end{document}